\documentclass{AIMS}
 \usepackage{amsmath}
 \usepackage{graphics} 

 \usepackage{hyperref} 

\usepackage{verbatim}

\def\currentvolume{X}
 \def\currentissue{X}
  \def\currentyear{200X}
   \def\currentmonth{XX}
    \def\ppages{X--XX}

\newtheorem{theorem}{Theorem}[section]
\newtheorem{corollary}{Corollary}

\newtheorem{lemma}[theorem]{Lemma}
\newtheorem{proposition}{Proposition}

\theoremstyle{definition}

\newcommand{\enproof}{\hfill $\Box$ \vspace*{1ex}}

\newcommand{\ssi}{^{(i)}} 

\newcommand{\mymathbb}[1]{{\mathbb{#1}}} 
\newcommand{\mymathsf}[1]{{\mathsf{#1}}} 

\newcommand{\dmn}{q} 

\newcommand{\sA}{\mymathsf{A}}

\newcommand{\sB}{\mymathsf{B}}

\newcommand{\cC}{{\mathcal C}} 

\newcommand{\myF}{{\mymathbb{F}_{\dmn}}} 
\newcommand{\myFnoarg}{{\mymathbb{F}}}
\newcommand{\myFk}{{\mymathbb{F}_{\dmn^k}}}

\newcommand{\cL}{{\mathcal L}}

\newcommand{\cP}{\mathcal{P}}

\newcommand{\sP}{\mymathsf{P}}
\newcommand{\sQ}{\mymathsf{Q}}

\newcommand{\cT}{{\mathcal T}}

\newcommand{\sU}{\mymathsf{U}}
\newcommand{\sZ}{\mymathsf{Z}}
\newcommand{\cY}{\myF}  
\newcommand{\cYY}{\myF} 
\newcommand{\cYpower}[1]{\myFpower{#1}}  

\newcommand{\vep}{\varepsilon}
\renewcommand{\phi}{\varphi} 
\renewcommand{\subset}{\subseteq}
\renewcommand{\tilde}{\widetilde}

\renewcommand{\bar}{\overline}

\newcommand{\defeq}{\stackrel{\rm def}{=}}

\newcommand{\SINT}{\mymathbb{Z}}

\newcommand{\transp}{\mbox{}^{\rm t}}
\newcommand{\supp}{\mymathsf{supp}\,}

\newcommand{\crd}[1]{|#1|}

\newcommand{\dpr}[2]{#1 \cdot #2}

\newcommand{\rvX}{\mymathsf{X}}

\newcommand{\rvZ}{\sZ} 

\newcommand{\rvu}{\sU}

\newcommand{\rcl}{r_{\rm c}} 

\newcommand{\myFpower}[1]{\mymathbb{F}_{\dmn}^{#1}}

\newcommand{\Qpl}[1]{\sQ^+} 
\newcommand{\Qmi}[1]{\sQ^-}
\newcommand{\vart}{t} 

\newcommand{\Bp}[1]{G}

\newcommand{\zrv}{0_n}

\newcommand{\Cgood}{C}
\newcommand{\Jgood}{\tilde{J}}

\newcommand{\Jof}[1]{\prm(\Jgood)}

\newcommand{\tsp}{M}
\newcommand{\tsptwo}[2]{\tsp_{#1}(#2)}

\newcommand{\prm}{\pi}

\newcommand{\chooses}[2]{\Big( \begin{array}{c} #1 \\[-1ex] #2 \end{array} \Big)}

\newcommand{\kperp}{\perp} 
\newcommand{\embone}{\pi_1}
\newcommand{\embtwo}{\pi_2}
\newcommand{\embj}{\pi_j}

\newcommand{\intint}[2]{[#1,#2]_{\SINT}} 

\newcommand{\degF}{k} 

\newcommand{\myFkpower}[1]{\myFnoarg_{\dmn^\degF}^{#1}}

\newcommand{\subgrp}{\subset} 

\newcommand{\CSone}{C_1}
\newcommand{\CStwo}{C_2^{\perp}} 
\newcommand{\CStwp}{C_2}
\newcommand{\Noa}{N_{\rm o}}
\newcommand{\Koa}{K_{\rm o}}
\newcommand{\Roa}{R_{\rm o}}

\newcommand{\Pej}{P_{{\rm e},j}}
\newcommand{\Peinj}{P_j}

\newcommand{\inu}{\nu}
\newcommand{\dblith}{} 
\newcommand{\Kdblith}[1]{K_{#1}} 
\newcommand{\ith}{_{\inu}} 

\newcommand{\varzd}{z}
\newcommand{\vardN}{z'}
\newcommand{\agd}{A}
\newcommand{\chP}{W}
\newcommand{\kgn}{\kappa}
\newcommand{\Rgn}{r}

\newcommand{\Nith}{N} 
\newcommand{\gith}{g} 
\newcommand{\dvsrith}{A} 
\newcommand{\divGith}{G} 
\newcommand{\vardNith}{z'} 

\title[Constructive Conjugate Codes]%
{Constructive Conjugate Codes for Quantum Error Correction and Cryptography}

\author{Mitsuru Hamada}


\email{mitsuru@ieee.org}

\thanks{This work was presented in part at 2006 IEEE Information Theory
Workshop, Chengdu, China, Oct., 2006.}

\begin{document}

\maketitle
\vspace*{-2ex}
{\footnotesize
 \centerline{Quantum Information Science Research Center}
 \centerline{Tamagawa University Research Institute}
 \centerline{6-1-1 Tamagawa-gakuen, Machida, Tokyo 194-8610, Japan \vspace*{.5ex}}
 \centerline{PRESTO, Japan Science and Technology Agency} 
 \centerline{4-1-8 Honcho, Kawaguchi, Saitama, Japan}
}

\medskip

 \centerline{(Communicated by an editor)}
 \medskip

\begin{abstract}
A conjugate code pair is defined as a pair of linear codes
either of which contains the dual of the other.
A conjugate code pair represents the essential structure
of the corresponding Calderbank-Shor-Steane (CSS) 
quantum error-correcting code.
It is known that conjugate code pairs are applicable to
quantum cryptography.
In this work, a polynomial construction of
conjugate code pairs is presented. 
The constructed pairs achieve the 
highest known achievable rate on additive channels,
and are decodable with algorithms of polynomial complexity.
\end{abstract}

\section{Introduction\label{ss:intro}}

Algebraic coding has proved useful not only on `classical' channels,
already in practical use,
but also on `quantum' channels, i.e., on
those that behave in quantum theoretical manners.
In particular, problems of secure information transmission through
quantum channels in the presence of eavesdroppers
have attracted great attention~\cite{BennettBrassard84,mayers01acm,ShorPreskill00} since a quantum key distribution protocol was proved secure
without resort to unproven computational assumptions~\cite{mayers01acm}.

On the one hand, 
there is a large literature of physics on design and analysis
of physical layers of such systems.
On the other hand, 
the design issue of codes in this context has left much room for 
investigation
whereas the influential work~\cite{ShorPreskill00}
on quantum key distribution suggests that the key technique 
of such cryptographic systems is algebraic coding.

Codes that can be used in cryptographic protocols, 
as well as for quantum error correction, 
have been treated in \cite{hamada03s,hamada05qc,hamada06s,hamada06ccc},
and the present work is a continuation.
The codes of our concern here
are essentially Calderbank-Shor-Steane (CSS) quantum codes
as in \cite{ShorPreskill00,hamada03s,hamada06s,hamada06ccc}. 
For CSS codes, however, we use an almost
synonym {\em conjugate code pairs},
which would represent the essence of CSS quantum codes. 
A conjugate code pair is defined as a pair of linear codes
either of which contains the dual of the other.

This work presents an `explicit', or more precisely, polynomial
construction of conjugate code pairs. 
The constructed codes fall in the class of concatenated conjugate codes
recently proposed by the present author~\cite{hamada06ccc}.
These are conjugate code pairs 
dually equipped with the structure of concatenated codes~\cite{forney}.

Recall there exist two major design criteria for error-correcting codes,
or simply codes:
(i) decoding error probability of a code,
and (ii) minimum distance of a code.
Whereas (ii) is adopted as a measure of the goodness of a code in 
\cite{hamada06md} to show that the codes in \cite{hamada06ccc}
are better than those known,
we adopt (i) in this paper. This would be legitimate
in the context of Shannon theory.

The remaining part of the paper is outlined here.
After preliminaries are given in Section~\ref{ss:pre},
the problem to be treated is described in Section~\ref{ss:cc}.
In Section~\ref{ss:app_good_spectrum},
a `balanced' ensemble of conjugate codes is given, and
then, it is argued that a good code exists in a balanced ensemble
of codes. 
In fact, a large portion of a balanced 
ensemble is made of good codes.
This fact is used in a polynomial construction of
codes in Section~\ref{ss:constructible}
while Sections~\ref{ss:qc_cc} and \ref{ss:ccc} are devoted to explaining
needed basic notions, 
i.e., quotient codes and concatenated conjugate codes, respectively.
We conclude with a summary and remarks in Section~\ref{ss:sum_rem}.

\section{Preliminaries\label{ss:pre}}

To focus on the design issue of codes, we begin directly with 
treating codes over a finite field, and evaluate the 
performance of codes in terms of classical probabilities.
How the evaluation implies the reliability 
and security of the corresponding quantum codes or
cryptographic protocols can be found in \cite{ShorPreskill00,hamada03s,hamada06s}.

We fix our notation, and recall some useful tools here.
As usual, $\lfloor a \rfloor$ denotes the largest
integer $a'$ with $a'\le a$, and $\lceil a \rceil = - \lfloor - a \rfloor$.
An $[n,k]$ linear (error-correcting) code over a finite field $\myF$, the finite field of $\dmn$
elements, is a $k$-dimensional subspace of $\myFpower{n}$.
The dual of a linear code $C \subset \myFpower{n}$ is
$\{ y \in\myFpower{n} \mid \forall x\in C, \ \dpr{x}{y}=0 \}$
and denoted by $C^{\perp}$,
where $\cdot$ denotes the ordinary (Euclidean) inner product.
The method of {\em types}\/ in information theory~\cite{csiszar_koerner,csiszar98} is a standard tool for elementary and combinatorial analysis
on probabilities.
Geometric Goppa codes (algebraic geometry codes) 
will be used in Section~\ref{ss:constructible},
while the same achievable rate can be obtained
with the more elementary Reed-Solomon (RS) codes
with a slower speed of convergence of error probability to zero~\cite{hamada06itw}.
Whereas geometric Goppa codes
were originally described in the language
of algebraic geometry, these can, more directly, be
described in the language of the function fields~\cite{stichtenoth}, 
which we will use.
The notion of polynomial constructibility is well established
in coding theory~\cite[p.~80]{TsfasmanVladut}, \cite[p.~317]{Stepanov}:
A sequence of codes $\{ C\ith \}$ of growing length
is said to be polynomially constructible or {\em polynomial}\/
if their encoders or generator matrices of $C\ith$
are constructible with polynomial complexity in their code-length.
Similarly, our conjugate codes to be presented are polynomial
in that generator matrices of them can be produced
with polynomial complexity as the conjugate codes in \cite{hamada06md}.

\section{Theme\label{ss:cc}}  

Consider a pair of linear codes $(C_1,C_2)$ satisfying
\begin{equation}\label{eq:css_cond}
\CStwo \subgrp \CSone, 
\end{equation}
which condition is equivalent to $\CSone^{\perp} \subgrp \CStwp$.
The following question arises from
an issue on quantum cryptography~\cite{ShorPreskill00} (also
explained in \cite{hamada06s,hamada05qc}):
How good
both $C_1$ and $C_2$ can be under the constraint
(\ref{eq:css_cond})?

We have named a pair $(C_1,C_2)$ with (\ref{eq:css_cond})
a conjugate (complementary) code pair, or more loosely, conjugate codes. 
We remark
that a CSS quantum code is specified by a conjugate code pair $(C_1,C_2)$
(put $\cC_1=C_1,\cC_2=C_2^{\perp}$ in \cite{CalderbankShor96}; $C_1$ corrects
bit flip errors and $C_2$ corrects phase errors). 
CSS codes form a class of algebraic quantum error-correcting codes called
symplectic or stabilizer codes (\cite{crss98} and references therein).

\section{Good Codes in a Balanced Ensemble\label{ss:app_good_spectrum}}

We can find good codes in an ensemble if the ensemble is balanced
in the following sense (e.g., \cite{hamada05qc}).
Suppose $\sA$ is an ensemble of subsets of $\myFpower{n}$.
If there exists a constant $V$ such that
$\crd{\{ \Cgood \in\sA \mid x\in \Cgood \}}=V$ for
any word $x \in \myFpower{n}\setminus\{ \zrv \}$, where
$\zrv\in\myFpower{n}$ is the zero vector,
the ensemble $\sA$ is said to be {\em balanced}.

We will construct
a relatively small ensemble $\sB$ of conjugate code pairs $(C_1,C_2)$ such that
both $\{ C_1 \mid (C_1,C_2) \in \sB \}$
and $\{ C_2 \mid (C_1,C_2) \in \sB \}$ are balanced.
Let $T$ be the companion matrix~\cite{LidlNied} 
(the definition can also be found in \cite{hamada06ccc}), or its transpose,
of a primitive 
polynomial of degree $n$ over $\myF$.
Given an $n\times n$ matrix $M$, let $M|^{m}$ (respectively, $M|_m$) denote the $m\times n$ submatrix
of $M$ that consists of the first (respectively, last) $m$ rows of $M$.
We put $C_1^{(i)}=\{ x T^i|^{k_1} \mid x\in\myFpower{k_1} \}$ and
$C_2^{(i)}=\{ x (T^{-i})\transp|_{k_2} \mid x\in\myFpower{k_2} \}$
for $i=0,\dots,\dmn^{n}-2$, where $M\transp$ denotes the transpose of $M$.
Then, setting
\begin{equation}\label{eq:ensT}
\sB=\sB(T)=\{ (C_1^{(i)},C_2^{(i)}) \mid i =0,\dots,\dmn^n-2 \},
\end{equation}
we have the next lemma. 

\begin{lemma}\label{lem:ens0}
For 
integers $k_1,k_2$ with $0 \le n-k_2 \le k_1 \le n$
and $\sB=\sB(T)$ constructed as above, any $(C_1,C_2)\in\sB$ is
a conjugate code pair, and both $\{ C_1 \mid (C_1,C_2) \in \sB \}$
and $\{ C_2 \mid (C_1,C_2) \in \sB \}$ are balanced.
\end{lemma}

{\em Remark.}\/
The matrix $T$ may, 
more generally, be an $n\times n$ invertible matrix over $\myF$ 
of the following properties.
(i) The set $\{ O_{n}, I_{n}, T, \ldots, T^{\dmn^{n} -2} \}$ 
is closed under the addition and multiplication,
where $O_n$ and $I_n$ denote the zero matrix and identity matrix, respectively.
(ii) The matrix $T$ has multiplicative order $\dmn^n-1$.

\begin{figure} 
\begin{center}
\scalebox{0.85}{\includegraphics
{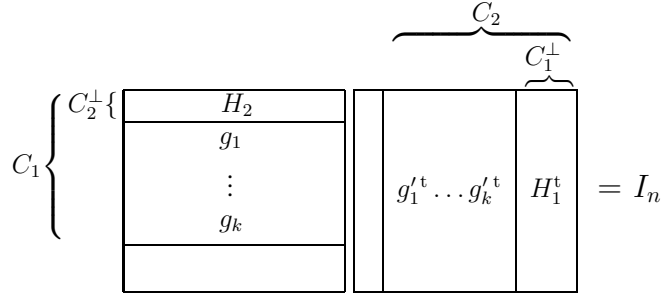}} 
\caption{%
A basic structure of an $[[n,k]]$ conjugate code pair.
 \label{fig:css}}
\end{center}
\end{figure}

{\em Proof of Lemma~\ref{lem:ens0}.}\/
The condition (\ref{eq:css_cond}) is fulfilled
since $T^i T^{-i}=I_n$ implies that
the $C_2\ssi\mbox{}^{\perp}$ is spanned by the first $n-k_2$ rows of $T^i$.
(Consider the situation where the matrix equation in Figure~\ref{fig:css}
is $T^i T^{-i}=I_n$. The picture was drawn 
for a generic conjugate code pair in \cite{hamada06ccc}.)
%

We can write 
$C_1^{(i)}=\{ y T^i \mid y\in\myFpower{n},\, 
\supp y \subset \intint{1}{k_1} \}$,
where $\supp (y_1,\dots,y_n)$ $=$ $\{ i \mid y_i \ne 0 \}$, 
and 
$\intint{i}{j}=\{ i,i+1,\ldots,j \}$ for integers $i \le j$.
Imagine we list up all codewords in $C_1^{(i)}$ allowing duplication.
Specifically, we list up all $y T^i$
as $y$ and $i$ vary over the range
$\{ y  \mid y\in\myFpower{n},\, \supp y \subset \intint{1}{k_1} \}$
and over $\intint{0}{\dmn^n-2}$, respectively.

With $y\in\myFpower{n}\setminus \{ 0 \}$ fixed, 
$y T^i$, $i\in\intint{0}{\dmn^{n-2}}$, are all distinct
since $T^i \ne T^j$ implies $y T^i - y T^{j}=y T^{l}$ for some 
$l$ and $y T^{l}$ is not zero.
Hence, any nonzero fixed word in $\myFpower{n}$ appears
exactly $\dmn^{k_1}-1$ times in listing $yT^i$ as above.
Namely, the ensemble $\{ C_1 \mid (C_1,C_2)\in \sB \}$ is balanced.
Using $(T^{-i})\transp$ in place of $T^i$, we see
the ensemble of $C_2$ is also balanced, completing the proof.
\enproof

To proceed, we need some notation.
We denote the type of $x\in\myFpower{n}$ by $\sP_x$~\cite{csiszar_koerner,csiszar98}.
This means that the number of appearances of $u\in\myF$ in $x\in\myFpower{n}$ is $n\sP_x(u)$.
The set of all types of sequences in $\cYpower{n}$
is denoted by $\cP_n(\cY)$.
Given a set $C\subset\cYpower{n}$, we put 
$\tsptwo{Q}{C}= \crd{\{ y\in C \mid \sP_y = Q \}}$
for types $Q\in\cP_n(\cY)$. 
The list of numbers $(\tsptwo{Q}{C})_{Q\in\cP_n(\cY)}$ 
may be called the $\sP$-spectrum (or simply, spectrum) of $C$.
For a 
type $Q$,
we put $\cT_{Q}^n=\{ y \in\cYpower{n} \mid \sP_{y}=Q \}$. 
We denote by $\cP(\cYY)$ the set of all probability distributions
on $\cYY$. 

We use the next proposition,
which relates the spectrum of a code
with its decoding error probability when it is used
on the additive memoryless channel characterized by a probability 
distribution $W$ on $\myF$, 
which changes an input $x\in\myF$ into $y$ with probability $W(y-x)$.
This kind of error bound has appeared in \cite{gallager63} while
the present form is from \cite[Theorem~4]{hamada05qc}.

\begin{proposition}\label{th:rcex}
Suppose we have an $[n,\kgn]$ linear code $C$ over $\myF$ such that 
\[
\tsptwo{Q}{C\setminus \{ 0_n \}}
\le a_n \dmn^{\kgn-n} \crd{\cT_{Q}^n},
 \quad Q\in\cP_n(\cYY)
\]
where $0_n$ is the zero vector in $\myFpower{n}$.
Then, if $a_n \ge 1$,
its decoding error probability with the minimum entropy syndrome decoding
is upper-bounded by 
\[
a_n \crd{\cP_n(\cYY)}^2 d^{-n E_{\rm r}(\chP,\Rgn)}
\]
for any additive channel $\chP$ of input-output alphabet\/ $\myF$,
where $\Rgn=\kgn/n$ and $E_{\rm r}(\chP,\Rgn)$ is the random coding exponent of 
$\chP$
defined by
\[
E_{\rm r}(\chP,\Rgn) = \min_{Q\in\cP(\cY)} [D(Q||\chP) + |1-\Rgn-H(Q)|^+].
\]
Here, $D$ and $H$ denote the relative entropy and entropy, respectively,
and $|x|^+ =\max\{ 0, x \}$.
\end{proposition}

In the simplest case where $\dmn=2$, the premise of the above proposition
reads `the spectrum
of $C$ is approximated by the binomial coefficients $\crd{\cT_{Q}^n}$ 
up to normalization.'
By a straightforward argument using Markov's inequality,
we have the next lemma
and estimate for the number of bad codes in (\ref{eq:frac_bad}) below
(a proof can be found in \cite{hamada06itw}).
\begin{lemma}\label{lem:C1C2spectrum}
We have
\begin{equation*}
\!\!\!
\tsptwo{Q}{C_j \setminus \{\zrv \} }
\le 2 (\crd{\cP_n(\myF)}-1) \dmn^{k_j-n} \crd{\cT_{Q}^n},\quad 
 Q\in\cP_n(\myF), \ j=1,2
\end{equation*}
for some conjugate code pair $(C_1,C_2)\in\sB$.
\end{lemma}

The codes $C_1$ and $C_2$ in Lemma~\ref{lem:C1C2spectrum}
both attain the random coding exponent by Proposition~\ref{th:rcex}.
Thus, we have a conjugate code pair $(C_1,C_2)$ such that 
both the $[n,k_1]$ code $C_1$ and the $[n,k_2]$ code $C_2$,
used on $\chP_1$ and on $\chP_2$,
have asymptotically vanishing decoding error probabilities,
as far as the asymptotic rates are below $1-H(\chP_1)$ and $1-H(\chP_2)$,
respectively.

In fact, a large portion of a balanced 
ensemble is made of good codes.
Specifically, let us say an $[n,k_j]$ code
$C_j\ssi$ is $\agd$-good if 
\begin{equation}\label{eq:a-good}
\tsptwo{Q}{C_j\ssi \setminus \{ 0_n \}} \le (\crd{\cP_n(\myF)}-1) q^{-n(1-r_j)} \agd
\end{equation}
for all $Q\in\cP_n(\myF)$, where $r_j = k_j/n$.
Then, by Markov's inequality, the number of codes that are not $\dmn^{\vep n}$-good
in $\{ C_j \mid (C_1,C_2) \in \sB \}$ is at most 
\begin{equation}\label{eq:frac_bad}
z=\lfloor N \dmn^{-\vep n} \rfloor.
\end{equation}
For $\dmn^{\vep n}$-good codes $C_j\ssi$, the decoding error probability
is upper-bounded by
\begin{equation}\label{eq:bound_inner}
\Peinj = a_n q^{-n[E_{\rm r}(\chP,r_j)-\vep]}, \quad j=1,2,
\end{equation}
where $a_n = \crd{\cP_n(\myF)}^3$ 
is at most polynomial in $n$,
owing to Proposition~\ref{th:rcex}.

\section{Quotient Codes and a Closer Look at Theme\label{ss:qc_cc}}

We have been seeking a pair $(C_1,C_2)$ with
(\ref{eq:css_cond}) such that both $C_1$ and $C_2$ are good.
However, this requirement can be weakened slightly.
In fact, we are interested in a pair $(C_1,C_2)$ such that
both quotient codes $C_1/C_2^{\perp}$ and $C_2/C_1^{\perp}$ are
good in the context of cryptography~\cite{ShorPreskill00,hamada06s,hamada05qc}
as well as in that of quantum error correction.

A {\em quotient code}\/ $C/B$ means an
additive quotient group $C/B$, and can be used for transmission
of information in the following way.  
We encode a message into a member of $c\in C/B$, which has the form
$x+B$, and then
transmit a randomly chosen word in $c$.
Clearly, if $C$ is $J$-correcting in the ordinary sense,
$C/B$ is $(J+B)$-correcting. For the purpose of transmission
of information only, dividing the code $C$ by $B$ does not seem meaningful,
but this scheme has proved useful for transmission 
of secret information in the presence of eavesdroppers.
The role of $B$ may be understood as a kind of scrambler. 
This scenario is due to \cite{wyner75}
while the term quotient code was coined in \cite{hamada05qc}.

As mathematical objects, the conjugate code pairs $(C_1, C_2)$ and the quotient codes $C/B$
are obviously in one-to-one correspondence: 
$(C_1, C_2) \leftrightarrow C_1/C_2^{\perp}$, or
$(C_1, C_2) \leftrightarrow C_2/C_1^{\perp}$.
We mean by an $[[n,k]]$ conjugate code pair
over $\myF$
a pair $(C_1,C_2)$ consisting of
an $[n,k_1]$ linear code $\CSone$ and an
$[n,k_2]$ linear code $C_2$ over $\myF$
satisfying (\ref{eq:css_cond})
and
\begin{equation}\label{eq:css_k}
k=k_1+k_2-n.
\end{equation}
The number $k/n$ is called the {\em (information) rate}\/ of 
the conjugate code pair $(C_1,C_2)$, and equals
that of quotient code $C_1/C_2^{\perp}$, that of $C_2/C_1^{\perp}$,
and that of the corresponding CSS quantum code.

The goal
is to find a conjugate code pair $(C_1,C_2)$ such that
both $C_1/C_2^{\perp}$ and $C_2/C_1^{\perp}$ 
are good 
on either of the criteria mentioned in Section~\ref{ss:intro}.
If the linear codes $C_1$ and $C_2$ are both good, so are $C_1/C_2^{\perp}$ and $C_2/C_1^{\perp}$. 
Hence, a conjugate code pair $(C_1,C_2)$ with good 
$C_1$ and $C_2$ is also desirable.

\section{Concatenated Conjugate Codes\label{ss:ccc}}

We recall how the {\em concatenation}\/ $(L_1,L_2)$
of inner $[[n\ssi,k]]$ conjugate code pairs
$(C_1\ssi,C_2\ssi)$ over $\myF$, $i=1,\dots,N$, 
and an outer $[[N,K]]$ conjugate code pair
$(D_1, D_2)$ over $\myFk$ was obtained in \cite{hamada06ccc} (the result also appeared in \cite{hamada06itw})
retaining the notation of \cite{hamada06ccc,hamada06itw}.
For $j=1,2$, let $\embj\ssi$ be a one-to-one $\myF$-linear map from $\myFk$
onto a set of coset representatives of
$C_j\ssi/C_{\bar{j}}\ssi\mbox{}^{\perp}$,
where $\bar{1}=2$ and $\bar{2}=1$,
and $\embj(x)$ denote the juxtaposition 
$\embj^{(1)}(x_1)\cdots \embj^{(N)}(x_N) \in \myFpower{\sum_i n\ssi}$
of $\embj^{(i)}(x_i)$, $i\in \{ 1,\dots,N \}$, 
for $x=(x_1,\dots,x_N)\in\myFkpower{N}$.
Then,
$L_j=\embj(D_j)+\overline{C_{\bar{j}}^{\perp}}$, 
where $\overline{C_{\bar{j}}^{\perp}}=\bigoplus_{i=1}^N 
C_{\bar{j}}\ssi\mbox{}^{\perp}$.

For $(L_1,L_2)$ to satisfy the constraint $L_2^{\perp} \subset L_1$,
we need a certain condition on $\embone$ and $\embtwo$.
In fact, by a proper choice of $\embone$ and $\embtwo$ based on
dual bases of $\myFk$ and the structure of conjugate codes as depicted in Figure~\ref{fig:css},
we have the next theorem.

\begin{theorem}\label{th:duals_ccc} \cite{hamada06ccc}
\[
\mbox{}\!\!\!\!\! \!\!\!\!\! \!\!\!\!\!
[\embone(D_2^{\kperp})+\overline{C_2^{\perp}}]^{\perp} = \embtwo(D_2)+\overline{C_1^{\perp}},
\quad
[\embtwo(D_1^{\kperp})+\overline{C_1^{\perp}}]^{\perp} = \embone(D_1)+\overline{C_2^{\perp}}.
\]
\mbox{}
\end{theorem}

\section{Polynomial Construction of Codes\label{ss:constructible}}

Using almost all codes in $\sB(T)$ 
as inner codes, we will construct the desired codes.
The point of the argument below is that, 
by Markov's inequality, all but a negligible
number of the inner codes are good (Section~\ref{ss:app_good_spectrum}),
so that the overall performance
is also good~\cite{DelsartePiret82}.

\begin{theorem}\label{th:main}
Let $\Roa$ be a number that can be written as
$\Roa=(r_1+r_2-1)(R_1+R_2-1)$ with some
$r_1,r_2,R_1,R_2\in (0,1]$, $r_1+r_2-1 \ge 0, R_1+R_2-1 \ge 0$.
Then, we have a sequence of $[[\Noa,\Koa ]]$ conjugate code pairs $(L_1,L_2)$
over $\myF$
such that the rate $\Koa/\Noa$ approaches $\Roa$, 
a parity-check matrix of $L_2$ and that of $L_1^{\perp}$ can
be produced with algorithms of polynomial complexity, 
and the decoding error probability $\Pej$
of $L_j/L_{\bar{j}}^{\perp}$,
where $\bar{1}=2$ and $\bar{2}=1$,
is bounded by 
\begin{equation*}
\limsup_{\Noa\to\infty} - \frac{1}{\Noa} \log_q \Pej 
\ge \frac{1}{2}(1-R_j)E(\chP_j,r_j), \quad  j=1,2
\end{equation*}
for any additive channels $\chP_1,\chP_2$ 
of input-output alphabet\/ $\myF$, where
$E(\chP_j,r_j)$ is the random coding exponent $E_{\rm r}(\chP_j,r_j)$.
Moreover, we have a polynomial decoding algorithm for
the quotient code $L_2/L_1^{\perp}$.
\end{theorem}

\begin{corollary}\label{coro:main}
Let a number $\Roa \in (0,1]$ be given.
Then, we have a sequence of conjugate code pairs $(L_1,L_2)$ satisfying
the same conditions as in the theorem except the bound on $\Pej$,
which is to be replaced by 
\[
\limsup_{\Noa\to\infty} - \frac{1}{\Noa} \log_q \Pej 
\ge \frac{1}{2}\sup_{r_1,r_2,R_1,R_2} \min_{l\in\{1,2\}}(1-R_l)E(\chP_l,r_l),
 \quad  j=1,2
\]
for any additive channels $\chP_1,\chP_2$,
where the supremum is taken over 
$\{ (r_1,r_2,R_1$, $R_2) \in (0,1]^4 \mid 
r_1+r_2-1\ge 0, \,  R_1+R_2-1 \ge 0, \,
(r_1+r_2-1)(R_1+R_2-1) = \Roa \}$.
\end{corollary}

{\em Remarks.}\/
The decoder for $L_2/L_1^{\perp}$, as well as the construction of 
$L_2/L_1^{\perp}$,
is independent from the channel.
The bound in the theorem is also valid for 
$E(\chP,\rcl)=\max\{ E_{\rm r}(\chP,\rcl), E_{\rm ex}(\chP,\rcl) \}$,
where $E_{\rm ex}$ is as in \cite[Theorem~4]{hamada05qc},
but in this case, 
the decoder for $L_2/L_1^{\perp}$ would depend on the channel in general.
Here, the decoding complexity should be understood as that of quotient codes
as described in Section~\ref{ss:qc_cc} or \cite{hamada05qc}, not that of quantum codes.

{\em Proof.}\/
We will construct $(L_1,L_2)$  by concatenation retaining the notation of
Section~\ref{ss:ccc}. 
We use almost all $(C_1\ssi,C_2\ssi)$ in $\sB$ for inner codes, where
$C_j\ssi$ is an $[n,k_j]$ code for all $i$ ($j=1,2$). 
For outer codes, we use
polynomially constructible
geometric Goppa codes of large minimum distance.
Namely, we use codes over $\myFk$,
where $q^k=p^m$ with some $p$ prime and $m$ even,
obtained from function fields
of many rational places (places of degree one). 
Specifically, we use a family of
function fields $F\ith/\myFk$, $\inu=1,2,\ldots$, 
such that $F\ith/\myFk$ has genus $\gith=g_{\inu,k}$ and at least
$\Nith+1=N_{\inu,k}+1$ rational places, and assume they satisfy
\begin{equation}\label{eq:cond_g}
\lim_{k\to\infty} \frac{\gith}{\Nith} =0
\end{equation}
for any $\inu$.
This is fulfilled, e.g., by
the second Garcia-Stichtenoth tower of function fields~\cite{GarciaStichtenoth96AS2} with $\gith \le q^{k\inu/2}$ and $\Nith$ as in (\ref{eq:cond_N}) below.
(We remark
unlike typical situations where we are concerned with
the limit of $\gith/\Nith$ as $\inu\to\infty$, 
we fix $\nu$ here in taking the limit. 
The level $\nu$ is to be fitted to the target rate of inner codes by
(\ref{eq:tune_nu}) below.)

If we put $\dvsrith=P_1+\cdots +P_{\Nith}$, where
$P_i$ are distinct rational places in $F\ith/\myFk$
and $\divGith$ is a divisor of $F\ith/\myFk$ such that
$\supp \divGith \cap \supp \dvsrith = \emptyset$,
we have a geometric Goppa code 
$C_{\cL}(\dvsrith,G)$ defined by
\[
C_{\cL}(\dvsrith,G) = \{ (f(P_1),\ldots,f(P_{\Nith})) \mid f \in \cL(G) \}
\]
where $\cL(G)=\{ x \in F\ith \mid (x) \ge -G \}\cup\{ 0 \}$,
and $(x)$ denotes the 
(principal) divisor of $x$
(e.g., as in \cite[p.~16]{stichtenoth}).
We assume both $D_1$ and $D_2^{\perp}$ 
are obtained in this way from the function field $F\ith/\myFk$:
$D_1 = C_{\cL}(\dvsrith,G_1\dblith)$ and 
$D_2 = C_{\cL}(\dvsrith,G_2\dblith)^{\perp}$.
We require $G_2\dblith \le G_1\dblith$
so that the CSS condition $D_2^{\perp} \subgrp D_1$ is fulfilled.
We also assume $2\gith -2 < \deg G_j\dblith  < \Nith$ for $j=1,2$.
Then, the dimension of $D_1$ is
\[
\Kdblith{1}=\dim G_1\dblith  = \deg G_1\dblith -\gith+1,
\]
and that of $D_2$ is
\[
\Kdblith{2}=\Nith-\dim G_2\dblith  = \Nith-\deg G_2\dblith +\gith-1.
\]
The designed distance of $D_1$ is $\Nith-\deg G_1\dblith$,
and that of $D_2$ is $\deg G_2\dblith -2 \gith +2$.

We consider the asymptotic situation 
where for arbitrarily fixed $0\le r_j^*, R_j^* \le 1$,
\begin{equation}\label{eq:Rr_star}
r_j \defeq k_j/n \to r_j^* \quad 
\mbox{and} \quad 
R_j \defeq \Kdblith{j}/\Nith \to R_j^* , 
\quad j=1,2
\end{equation}
as $n,N \to \infty$,
using a family of $[\Nith,\Kdblith{j}]$ geometric Goppa codes 
that fulfills the following requirement  
as well as (\ref{eq:cond_g}).
Specific examples of such codes can be found in \cite{ShumAKSD01}
for $\inu \ge 3$
(we can use the RS and Hermitian codes for $\inu=1,2$)
or in \cite{Shen93}.
We assume that $\Nith$ grows fast enough so as to be almost as large as
the size of $\sB$ as $k\to\infty$.
Specifically, for any $\inu$, we require
\begin{equation}\label{eq:cond_N}
\Nith = q^{k(\nu+1)/2} - \vardNith
\end{equation}
for some $\vardNith=\vardN_{\inu,k}$
such that $\vardNith/\Nith\to 0$ as $k\to\infty$.
By this assumption, if we set 
\begin{equation}\label{eq:tune_nu}
\nu = \lceil 2n/k \rceil -1
\end{equation}
so that $q^{k(\nu+1)/2} > q^n -1 = \crd{\sB}$, 
we can use all but a negligible number $\le \vardNith$ 
of code pairs in $\sB$ as inner codes.

For decoding, 
we first decode the inner codes and then the outer code~\cite{hamada06ccc}.
Recall we have (\ref{eq:bound_inner}). 
Then, employing a decoder that can correct
$\lfloor (\Nith-\Kdblith{j})/2 \rfloor -\gith$ errors
for the outer codes, 
we have the decoding error probability $\Pej$ of $L_j/L_{\bar{j}}$
bounded by
\begin{eqnarray*}
\Pej &\le & \sum_{i=\vart-\varzd}^{\Nith-\varzd} \chooses{\Nith-\varzd}{i} \Peinj^i (1-\Peinj)^{\Nith-\varzd-i}\\
 &\le & q^{(\vart-\varzd) \log_{q} \Peinj + (\Nith-\vart) \log_{q}(1-\Peinj) + (\Nith-\varzd) h((\vart-\varzd)/(\Nith-\varzd)) }
\end{eqnarray*}
where $z$ is as in (\ref{eq:frac_bad}),
$h$ is the binary entropy function, 
and $\vart=\vart_j=\lfloor (\Nith-\Kdblith{j})/2 \rfloor -\gith +1$
(for the second inequality, see, e.g., \cite[p.~446]{roman}).
Taking logarithms and dividing by $\Noa=n\Nith$,
we have
\begin{eqnarray*}
\frac{1}{\Noa} \log_q \Pej &\le& 
\frac{\vart-\varzd}{\Nith} \Big[-E(W,r_j)+\vep+ \frac{\log_q a_n}{n} \Big]\\
&&\!\!\!\!\! \mbox{}+ \frac{1}{n}\Big[ \frac{\Nith-\vart}{\Nith} \log_q(1-\Peinj) + \frac{\Nith-\varzd}{\Nith} h\big((\vart-\varzd)/(\Nith-\varzd)\big)\Big]
\end{eqnarray*}
for $j=1,2$.
Hence, letting $k\to\infty$ 
with constraint (\ref{eq:Rr_star}) and using (\ref{eq:cond_g}),
we have
\begin{equation}\label{eq:exponential}
\limsup_{\Noa\to\infty} - \frac{1}{\Noa} \log_q \Pej 
\ge \frac{1}{2} (1-R_j^*) [E(W,r_j^*)-\vep].
\end{equation}
Then, since $\vep>0$ is arbitrary, 
noticing the overall rate is $kK/(nN)
=(r_1+r_2-1)(R_1+R_2-1)$,
we have the desired bounds in the theorem and corollary.

We have used 
a family of geometric Goppa codes $D_1$ and $D_2^{\perp}$
for which we have polynomial algorithms
to produce generator matrices~\cite{ShumAKSD01,ShumPhD},
and parity-check matrices and generator matrices 
of concatenated conjugate codes are easily obtained from those of $D_j$~\cite{hamada06ccc}.
In general,
if the outer code $D_j$ is decodable with polynomial complexity, 
so is the concatenated quotient code
$L_j/L_{\bar{j}}^{\perp}$~\cite{forney,hamada06ccc}.
For our choice~\cite{ShumAKSD01,Shen93}, 
$D_2=C_{\cL}(\dvsrith,G_2\dblith)^{\perp}$ is a `one-point' code, i.e., 
the support of $G_2\dblith$
consists of one rational place,
so that we have a polynomial decoding algorithm
that corrects
$\lfloor (\Nith-\Kdblith{2})/2 \rfloor -\gith$ errors
(e.g., \cite{stichtenoth},\cite{HvLPellikaan98}).
The proof is complete.
\enproof 

We remark that putting constraints $R_1^*=R_2^*$ and $r_1^*=r_2^*$
in (\ref{eq:exponential}), we have a weakened but closed form of the bound,
which appeared in \cite{hamada06itw}.

The highest achievable rate 
($\Roa$ such that $\Pej \to  0$ for both $j=1$ and $2$)
resulting from Theorem~\ref{th:main}
is $R_{\rm CSS}=1-H(W_1)-H(W_2)$ while
the bound in \cite{hamada06itw} mentioned above 
implies the weaker 
achievable rate
$R_{\rm CSS}'=1-2 \max\{ H(W_1),H(W_2) \}$.
The achievability of 
$R_{\rm CSS}'$ ($\le R_{\rm CSS}$)
by non-constructive conjugate (CSS) codes had been established
in \cite{hamada03s}
and by non-constructive but polynomially decodable conjugate codes 
in \cite{hamada06ccc}.

The improvement on $R_{\rm CSS}'$ is not novel. In fact,
a bound on the error probability in \cite[Section~10.3]{hamada05qc}
implies the achievability of $R_{\rm CSS}$
by non-constructive codes,
and the argument of \cite{hamada06ccc} can easily be amended
to establish the achievability of $R_{\rm CSS}$ using
a good inner code shown to exist in \cite[Section~10.3]{hamada05qc}.
Lemma~\ref{lem:C1C2spectrum} says we can find good codes
in an ensemble much smaller than
that of \cite[Section~10.3]{hamada05qc}.

We remark that in the extreme case where
$C_1\ssi=\myFpower{n}$ and $D_1=\myFkpower{\Nith}$,
the quotient code $L_2/L_1^{\perp}$ becomes the classical
concatenated code $L_2$,
and (\ref{eq:exponential}) applies even to this case. 

\section{Summary and Remarks\label{ss:sum_rem}} 

In this work, conjugate code pairs that are constructible
with polynomial complexity were presented. 
The constructed pairs achieve the 
highest known achievable rate on additive channels.
Moreover, the constructed codes, as quotient codes, 
allow decoding of polynomial complexity.
We conclude with some remarks.

(I) A polynomial construction of asymptotically good quantum codes 
was first presented in \cite{AshikhminLT01}. This adopts the 
criterion of minimum distance and reflects
the idea of polynomial constructions of classical algebraic codes
(\cite{TsfasmanVladut} and references therein). 
See \cite{hamada06md} for
attainable minimum distance of concatenated conjugate codes or quantum codes
of analogous structure, and how it is related to \cite{AshikhminLT01}.
The present work's approach is rather close to that of \cite{DelsartePiret82},
which presented an explicit construction of capacity-achieving codes
adopting the criterion of decoding error probability.

(II) Though a conjugate code pair is defined as
a pair of {\em classical}\/ codes,
it is rephrased as a CSS quantum error-correcting 
code (Section~\ref{ss:cc}).
Theorem~\ref{th:main} implies that our codes, as polynomially constructible quantum codes, 
achieve the rate $R_{\rm CSS}=1-H(P_{\rvX})-H(P_{\rvZ})$
for the channel that changes a quantum state $\rho$ into
$X^x Z^z \rho (X^xZ^z)^{-1}$ with probability $P_{\rvX\rvZ}(x,z)$, 
$(x,z)\in (\SINT/\dmn\SINT)^2=\{0,\dots,\dmn-1 \}^2$. 
Here, $\dmn$ is a prime, $X$ and $Z$ 
are unitary operators that represent unit shifts
in complementary observables
(e.g., they are distinct Pauli operators for $\dmn=2$),
and $P_{\rvu}$
denotes the distribution of a random variable $\rvu$.
(See, e.g., \cite{hamada03f,hamada05qc,hamada06s} for backgrounds.)
This holds true for general quantum channels
as well~\cite{hamada03f,hamada01g}.

(III) This work was motivated by
an issue on quantum cryptography, or specifically,
by the fact that good conjugate (CSS) codes 
can be used as the main ingredients of
some quantum cryptographic protocols
as argued in \cite{ShorPreskill00,hamada03s,hamada06s}.
The security of such protocols using our codes can be evaluated 
along the lines of \cite{ShorPreskill00,hamada03s}.

\end{document}